\documentstyle[aps,epsfig,amssymb,balanced,times,preprint]{revtex}
\tightenlines
\begin{document}
\draft
\title{Nonlinear Pattern Selection in Binary Mixture Convection with Through-Flow}
\author{Guo-Dong Li$^{1,3}$, Atsushi Ogawa$^2$, Yoshifumi Harada$^2$, and
Michael I.Tribelsky$^1$}
\address{$^1$Department of Applied Physics, $^2$Department of Human and
Artificial Intelligent Systems,\\ Faculty of Engineering, Fukui
University, Bunkyo 3-9-1, Fukui 910-8507, Japan\\ $^3$Institute for
Fluid Dynamics, Xi'An University of Technology, Xi'An 710048,
China}
\date{\today}
\maketitle
\begin{abstract}
The pattern selection problem in binary-mixture convection in an
extended channel with a lateral through-flow is presented. The
through-flow breaks left--right parity and changes pattern dynamics
dramatically. The problem is studied based on computer simulation
of the complete set of hydrodynamic equations (Oberbeck-Boussinesq
approximation) in the two-dimensional rectangular channel with
aspect ratio $\Gamma = 12$ and convection-suppressing lateral
boundary conditions. A wide variety of new dynamical patterns is
obtained, discussed and classified.\\
\end{abstract}

Convection in a horizontal layer of binary fluid mixture heated
from below is a paradigm for oscillatory pattern forming
systems\cite{CH}. An important generalization of the conventional
type of the problem is the convection with a lateral through-flow.
This generalization provides the unique opportunity to study
pattern formation in a system with broken left-right parity. The
symmetry breaking caused by the throw-flow lifts degeneracy of the
problem and for this reason must affect the pattern dynamics
dramatically. Thus, study of this type of convection sheds a light
on generic aspects of pattern formation in systems with broken
parity, which attaches especial importance to the problem.

The linear stability analysis (LSA) of the spatially uniform
conducting state performed in Ref.\cite{JLB} reveals that the
through-flow splits the single Hopf bifurcation point,
corresponding to the convection threshold without the trough-flow,
into two Hopf bifurcation points for upstream (UTW) and downstream
(DTW) traveling waves, respectively. It is important that the
values of the thresholds as well as their relative positions with
respect to each other depend on the value of the Reynolds number
({\it Re}) for the through-flow. Nonlinear effects are studied
numerically in recent publication\cite{LB}, where a number of
interesting results is obtained. However, the small value of aspect
ratio employed in Ref.\cite{LB} does not allow the authors to
answer the most appealing question "How does the trough-flow affect
nonlinear stage of pattern formation and pattern selection in an
extended channel, where the pattern structure is not imposed by the
lateral boundary conditions?"

The first attempt to answer this question is made in the present
Letter. To this end we perform numerical integration of the
standard set of two-dimensional (one horizontal coordinate $x$ and
vertical coordinate $z$, where $z$-axis is antiparallel to the
gravitational acceleration) hydrodynamic equations in the
Oberbeck-Boussinesq approximation. The equations may be found,
e.g., in Ref.\cite{LB} and are not written here explicitly. We
employ the usual non-dimensional variables, where length is
measured in units of the layer width $d$ and time in units of heat
diffusion time $d^2/\chi$ (here $\chi$ is the thermometric
conductivity).

The aspect ratio $\Gamma$ is chosen equal to 12. The boundary
conditions at the upper and lower confining surfaces are rigid,
isothermal and impermeable. The lateral boundary conditions for the
temperature and concentration fields at the inlet ($x=0$) yield the
basic linear profiles, while at the outlet ($x=\Gamma$) we require
vanishing of the corresponding $x$-derivatives. As for the lateral
boundary conditions for the velocity field, they impose the same
Poiseuille profile both at the inlet and outlet. These boundary
conditions simulate the ones, which may be imposed in real
experiment, cf., e.g., the boundary conditions for similar problems
with a through-flow employed in Refs.\cite{RB,TC}. In addition,
suppression of the convection close to the in- and outlet allows us
to minimize influence of the lateral boundaries on the pattern
selection process. The latter has especial importance due to
possibility of advection with the trough-flow perturbations
generated by the inlet into the bulk of the channel.

The Prandtl number ($Pr \equiv \nu/\chi$, where $\nu$ stands for
the kinematic viscosity) is chosen equal to 10, the Lewis number
($L \equiv D/\chi$, where $D$ is the concentration diffusion
coefficient) is 0.01. Such a choice of the constants corresponds to
commonly used in experiment room temperature ethanol-water
solution. For the separation ratio $\psi
\equiv S_TC_0(1-C_0)\beta/\alpha$, which is the measure of
coupling between the temperature and the concentration gradients
coursed by the Soret effect, the moderate value $\psi = -0.1$ is
chosen. Here $\alpha$ and $\beta$ are the thermal and solute
expansion coefficients, $S_T$ stands for the Soret coefficient and
$C_0$ is the mean concentration of the solution. The role of
control parameter plays the reduced Rayleigh number $r$, which at
the fixed values of the material constants of the fluid, actually,
equals $\Delta T/\Delta T_c$, where $\Delta T$ designates the
temperature difference between the bottom and the top of the layer
and $\Delta T_c$ corresponds to the threshold of instability of the
conductive state of the pure fluid without the through-flow $(Re =
0,\;\psi = 0)$ against infinitesimal perturbations. In simulations
discussed in the present Letter $r$ varies in rather a narrow
interval close to $r=1.2$. The trough-flow runs along the positive
direction of $x$-axis and remains the same in all the simulations.
The corresponding Reynolds number is 0.08\cite{n}. At these $r$ and
{\it Re} the characteristic values of the velocity of the shear
flow are about 15--20\% of the amplitude of the saturated
convection velocity, i.e., the through-flow is always weak.

The SIMPLE code\cite{code} with 242$\times$22 grid points and
temporal step $5\cdot10^{-4}$ is employed for the integration. The
simulation gives rise to the following results.

We begin with $r=1.20$. This value of $r$ lies close to the
thresholds of instability of the conducting state against
infinitesimal perturbations for both types of waves UTW and
DTW\cite{note}. The initial conditions are the conductive state
with the temperature field perturbed by random noise with the
amplitude about $10^{-4}$.

The simulations show that initially the conductive state becomes
unstable against both the types of the waves. The corresponding
pattern consists of superimposed, counter-propagating UTW and DTW.
At the discussed initial stage of the instability the profiles of
hydrodynamic variables for each type of the waves obey the
relations following from the LSA (we will name these waves {\it
linear} waves), being proportional to
\begin{equation}\label{LS}
  \exp[\gamma^{U,D} t + \mu^{U,D}x + i(k^{U,D}x - \omega^{U,D}t)],
\end{equation}
and similar to those in binary mixture without the
trough-flow\cite{Cross}. Here the superscripts $U$ and $D$ stand
for up- and downstream propagating waves and $\omega^U<0,\;
\mu^U<0,\; \omega^D>0,\;
\mu^D>0$. The wavenumbers for both types of the observed
waves $k^{U,D} \approx 3.2$, are quite close to the values of the
critical wavenumbers (the ones maximizing the real parts of the
growth rates for UTW and DTW at the corresponding thresholds),
given under the specified conditions by LSA\cite{note1}. As for
$\omega^{U,D}$, the expression obtained in Ref.\cite{JLB} says that
in the case under consideration for both types of the waves
\begin{equation}\label{omc}
  \omega^{U,D}_c \approx \pm 6.47 + 41.9 Re,
\end{equation}
where $\omega^{U,D}_c \equiv \omega^{U,D}(k_c)$ at the instability
threshold and signs plus and minus are related to down- and
upstream waves respectively. The two terms in r.h.s. of
Eq.~(\ref{omc}) have clear physical meaning, the first ($\pm 6.47$)
is associated with the Hopf bifurcation without the through-flow,
while the second describes the drift-effect caused by the
through-flow. The observed values of $\omega^{U,D}$ ($\omega^U
\approx -3.32,\; \omega^D
\approx 9.40$) also are in reasonable agreement with those given
by the above expression ($\omega^U_c \approx -3.11,\; \omega^D
\approx 9.82$).

However, the real part of the growth rate for UTW is bigger than
that for DTW\cite{JLB}. Accordingly, soon UTW overtake DTW and
begin to suppress them. During this process the amplitude
modulations caused by superposition of the two types of waves
become weaker and weaker. Finally, DTW are suppressed completely
and oscillatory growth of the amplitudes of convective modes is
transformed into purely exponential. In other words, the system
evolves to the state with the entire channel filled with UTW, who
still obey Eq.~(\ref{LS}). It happens at $t \approx 3.5$, when the
maximal amplitude of the $z$-component of the velocity field is
about $10^{-2}$. Thus, the first nonlinear effect observed in the
simulation is the suppression of DTW, which agrees with the results
obtained in Ref.\cite{LB} for small aspect ratio.

\begin{figure}
\vspace*{-4cm}
\epsfxsize = 1.1\textwidth
\makebox{\hspace{-2cm}\epsfbox{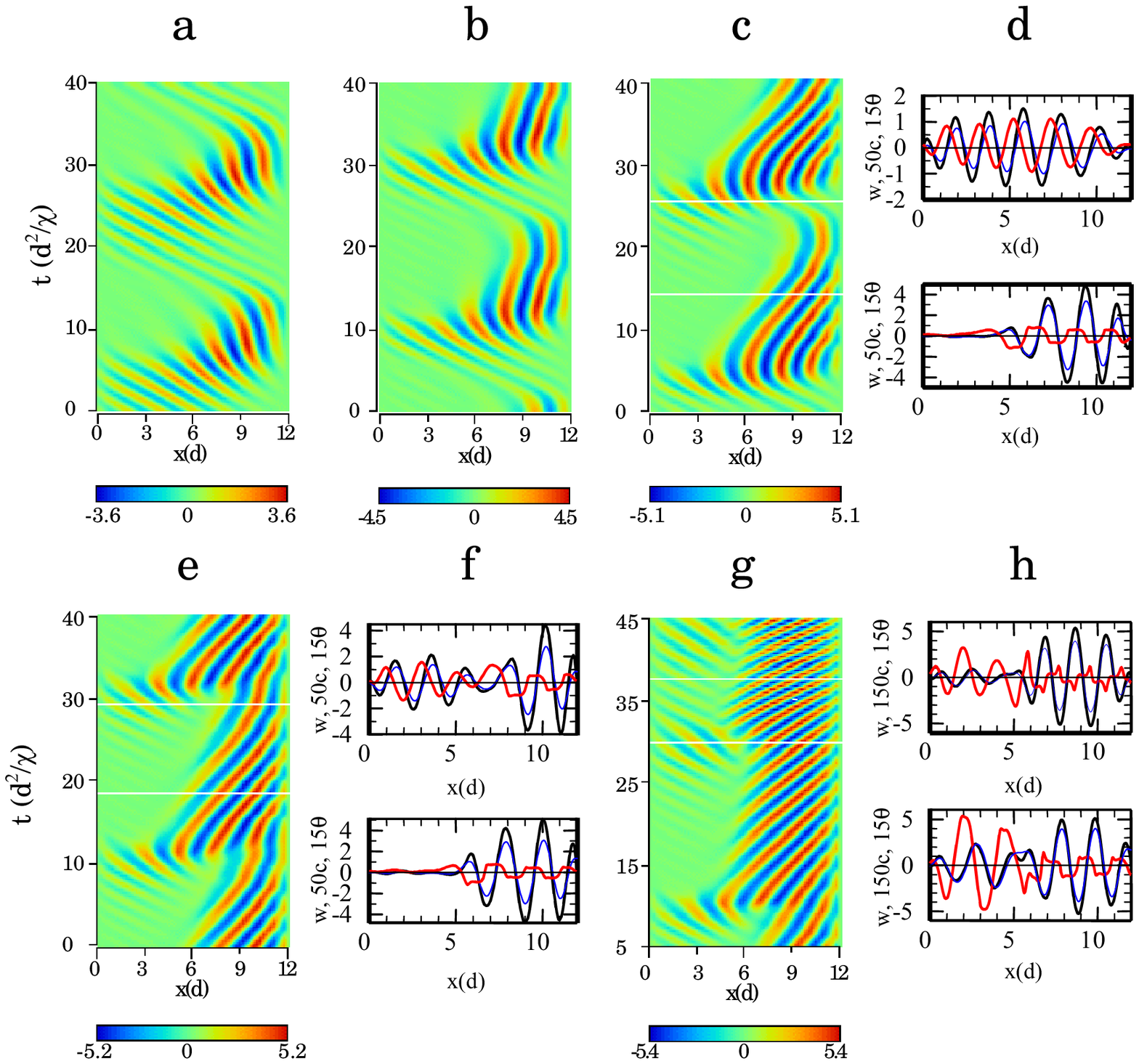}}
\vspace*{-6cm}
\caption{Spatiotemporal plots for $z$-component of
the velocity field in the midplane $w$ at different values of the
Rayleigh number and profiles of $w$ (black), concentration $c
\equiv (C-C_0)\beta/(\alpha \Delta T)$ (red) and
temperature $\theta \equiv (T-T_0)/\Delta T$ (blue) fields in the
midplane at some characteristic moments of time. a, b, c, e ---
asymptotic, periodically repeated in time patterns at $r=1.16$ (a),
$r=1.18$ (b), $r=1.20$ (c) and $r=1.21$ (e), respectively. d ---
profiles of $w$, $c$ and $\theta$ for spatiotemporal plot c at two
moments of time indicated in c as white lines, the upper and lower
panels on d correspond to the upper and lower lines on c. f ---
analogous profiles for e. g --- transition FDTW $\rightarrow$ DTW
initiated by abrupt raise of $r$ from 1.21 to 1.213 at $t=0$. h~---
profiles of $w$, $c$ and $\theta$ for plot g. The profiles in the
lower panel (the lower white line on g) correspond to FDTW, while
those in the upper panel (the upper white line on g) already to
DTW. For more details see the text.
\label{120}}
\end{figure}

Then, the growth of the amplitude of UTW brings about the next
nonlinear effects, which change the evolution of UTW themselves. As
usual\cite{CH}, when the amplitude becomes big enough nonlinear
stabilization occurs. The corresponding nonlinear terms affect both
the real and imaginary parts of the growth rate. Thus, when the
amplitude reaches the vicinity of its quasi-steady value,
determined by the balance between linear instability and nonlinear
stabilization, the frequency of UTW begins to increase sharply [the
frequency for the linear UTW is {\it negative}, see
Eq.~(\ref{omc}), so the increase of the frequency means decrease of
its absolute value]. It gives rise to slowing down of the phase
velocity of the waves and finally to their reversion, see
Fig.~\ref{120}c.

In the left half of the channel the scenario is different due to
influence of the convection-suppressing conditions drifted from the
inlet with the trough-flow. Here the exponential growth of the
waves' amplitude suddenly, before the amplitude achieves the
quasi-steady value, is changed with sharp decay. The convection in
this half of the channel is suppressed and the fluid returns
practically to the conductive state. The case is somewhat similar
to that for convection in pure fluid with through-flow, where
highly nonlinear states may be observed only at a certain distance
from the inlet\cite{pure1,pure2}.

Regarding the reversed waves, in the end of this stage they
transform into a quasi-steady wave train with pronounced left
boundary propagating to the outlet. However, actually these waves
remain UTW, which are just drifted downstream due to advection with
the trough-flow. The concentration mode for these waves is shifted
about one-quarter wavelength to the {\it left} of the $w$ mode,
where $w$ stands for $z$-component of the velocity field in the
midplane, so that counterclockwise circulating rolls have higher
concentration level than that for the clockwise, see
Fig.~\ref{120}d. Such a phase shift is typical to UTW, while for
DTW the phase shift has the opposite sign. Another evidence of the
UTW nature of these waves is their small phase velocity $v_{ph}
\approx 0.43$, which is smaller than both the maximal velocity of
the trough flow $v^{max}_{fl}=1.20$ and its average velocity
$\bar{v}_{fl}=0.8$.

At small Reynolds numbers we can write $v_{ph}(r,Re) \approx
v_{ph}(r,0) + v^\prime_{ph}(r,0)Re$, where the second term in the
r.h.s. of this expression may be regarded as the drift velocity of
the pattern $v_{dr}$. Generally speaking, the value of $v_{dr}$
should be different from both $v^{max}_{fl}$ and $\bar{v}_{fl}$ due
to the problem nonlinearity. To obtain $v_{dr}$ the following
approach is employed. For a given pattern we switch off the
trough-flow (replacing the lateral boundary conditions for the
velocity field with rigid, impermeable) and observe the dynamics of
the same pattern but without the through-flow.

As soon as the trough-flow is switched off the described
quasi-steady waves change the direction of propagation and begin to
travel upstream with the phase velocity $v_{ph} \approx -0.56$,
which yields $v_{dr}$ the value $v_{dr} \approx 0.99$ very close to
the drift velocity given by Eq.~(\ref{omc}) for the linear waves.
Bearing in mind all mentioned above, we will name these nonlinear
quasi-steady waves drifted downstream {\it false downstream
traveling waves} (FDTW).

Since DTW are suppressed, the reversion of the left marginal wave
in the UTW wave train creates on $xt$-plane a region shielded for
perturbations from past to enter. The conductive state in this
``restricted area" is destroyed with growth of small perturbations
generated by the boundary between this region and the one filled
with FDTW. It gives rise to secondary linear UTW propagating from
the boundary of the FDTW wave train to the inlet, whose profiles
are well described by Eq.~(\ref{LS}), see Fig.~\ref{120}c.

One by one the FDTW reach the outlet and vanish there. It results
in contraction of the domain filled with FDTW and expansion of the
one filled with linear UTW, respectively. It could be expected that
finally all FDTW disappear by the outlet but, actually, it is not
the case. Suddenly the amplitudes of FDTW begin to decay very fast.
The decay yields deceleration of FDTW in the laboratory coordinate
frame and eventually change the direction of their propagation,
i.e., the inverse transformation from FDTW to UTW occurs. The
entire channel becomes filled with linear UTW and then the cycle
(UTW $\rightarrow$ UTW + FDTW $\rightarrow$ UTW) is repeated
periodically with the period about 24 time units, see
Fig.~\ref{120}c. Possible qualitative explanation of this
re-linearization of the convection may be as follows. The narrower
the domain filled with FDTW the weaker its stability. When the left
boundary of the wave train advances close enough to the outlet, the
wave train cannot stand against convection-suppressing influence of
the lateral boundary conditions, which eventually results in
decrease of the waves' amplitude, i.e., to the re-linearization.

To examine how the described nonlinear evolution depends on the
Rayleigh number the simulations are repeated at different values of
$r$. However, under the specified conditions the bifurcation from
the conductive to convective state in binary mixture is
subcritical\cite{CH} and the dependence of pattern structure on the
Rayleigh number must exhibit a hysteretic loop. In this case to
avoid jumps to another branch of the bifurcation curve the patterns
obtained as the asymptotic state at the previous value of $r$ are
employed as the initial conditions for the new values of $r$.

\begin{figure}
\vspace*{3cm}
\epsfxsize = 2.0\textwidth
\makebox{\hspace{-7cm}\epsfbox{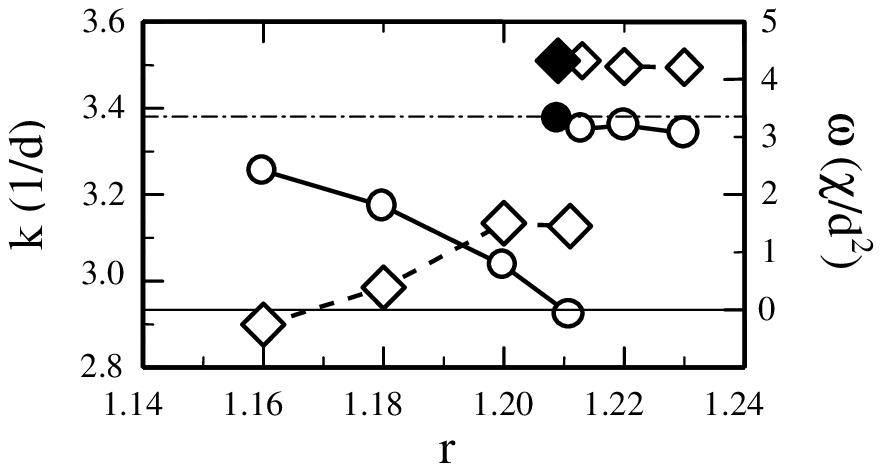}}
\vspace*{-32cm}
\caption{The dependence of the wavenumber $k$ ($\circ$) and oscillation
frequency $\omega$ ($\diamond$) on the reduced Rayleigh number $r$
for nonlinear quasi-steady UTW, FDTW and DTW. Note the
discontinuity at $r=1.213$ caused by transition FDTW $\rightarrow$
DTW. Closed symbols designate the boundary of a hysteretic loop
observed during decrease of $r.$ Full straight line in the plot
corresponds to $\omega =0$, dashed to the drift contribution of the
through-flow for the linear waves [second term in the r.h.s. of
Eq.~(\protect{\ref{omc}})].
\label{k}}
\end{figure}

The simulations show that in the region from the lowest inspected
value of $r=1.16$ up to $r=1.21$ the pattern formation basically
corresponds to the scenario discussed above. The saturated
amplitude of the quasi-steady waves and the corresponding frequency
both occur monotonically increasing functions of $r$, see
Figs.~\ref{120}--\ref{k}. The phase velocity of the quasi-steady
waves vanishes at $r \approx 1.17$. Accordingly, FDTW do not exist
at $r$ smaller than this value, cf. Fig.~\ref{120}a and b.

Since the effect of the sudden decay of convective modes definitely
is connected with the convection-suppressing lateral boundary
conditions, it becomes more pronounced with decrease of $r$
resulting in general degradation of the convection. In accordance
with that the width of the decay region increases, while the
fraction of the period, when the quasi-steady nonlinear waves
exist, decreases with decrease of $r$, cf. Fig.~\ref{120} c, b and
a. Alternatively increase of $r$ eventually eliminates the inverse
transition FDTW $\rightarrow$ UTW at all, see Fig.~\ref{120}e, f.

Increase of the Rayleigh number from $r=1.21$ to $r=1.213$ brings
about a new bifurcation. FDTW are accelerated. Both the amplitude
of FDTW and their phase velocity increase and finally the phase
velocity becomes equal to $v_{ph}\approx 1.20$, which is {\it
bigger} than the drift velocity, Fig.~\ref{120}g. Thus, FDTW
transforms into true DTW. The transformation is accompanied with
the corresponding change of the phase shift between the
concentration and $w$ modes, so that now counterclockwise
circulating rolls have smaller concentration level than that for
the clockwise, see Fig.~\ref{120}h. In the same time linear UTW in
the left half of the channel do not undergo the sudden decay any
more. Eventually the system evolves to a steady state with left
part of the channel filled with the linear UTW, which obey
Eq.~(\ref{LS}), and right part with the nonlinear steady DTW. The
boundary between these two domains remains practically immobile.

Another important feature of the problem revealed in simulations is
strict wavenumber and frequency selection. Every value of the
Rayleigh number yields a certain unique pair of $k$ and $\omega$
for each of the discussed types of the waves. Any change of $r$
triggers transient process, which results in the corresponding
changes of $k$ and $\omega$. The dependence of $k$ and $\omega$ on
$r$ for nonlinear quasi-steady UTW, FDTW and DTW is presented in
Fig.~\ref{k}. It is important to stress that the imposed boundary
conditions prevent any phase pinning or wavenumber discretization,
so the observed wavenumber selection is the generic, intrinsic
feature of the problem, which is not related to any external
factor.

Note that transition UTW $\rightarrow$ FDTW does not change the
smoothness of curves $k(r)$ and $\omega(r)$, while transition FDTW
$\rightarrow$ DTW does, see Fig.~\ref{k}. An interesting feature of
the phenomenon is that the functions $k(r)$ and $\omega(r)$ for UTW
and FDTW are much sharper than those for DTW, cf. the behavior of
these function blow and beyond the transition point. At decrease of
$r$ a narrow hysteretic loop for inverse transition DTW
$\rightarrow$ FDTW is observed. DTW remain stable until $r=2.09$.
Further decrease of the Rayleigh number to $r=2.08$ triggers
transformation DTW $\rightarrow$ FDTW with formation of a pattern
similar to that shown in Fig.~\ref{120}c.

The last issue to be discussed is the global stability of the
observed waves. According to the results of Ref.\cite{LB} for an
unbounded in the lateral dimension channel steady nonlinear UTW
should be unstable at employed in our simulations values of
$\psi=-0.1$, $Re=0.08$ and {\it any} value of $r$. The stable
patterns observed in our simulations do not contradict to this
result because their stability may be associated with the
size-effect at the moderate value of $\Gamma=12$ used in the
simulations. To elucidate this question simulations with bigger
value of $\Gamma$ are required. These simulations are under way and
will be reported elsewhere.

This work was supported by the Grant-in-Aid for Scientific Research
(No. 11837006) from the Ministry of Education, Culture, Sports,
Science and Technology (Japan). Japanese Government Scholarship for
one of us (G. D. Li) is gratefully acknowledged.

\begin{references}
\bibitem{CH}M.C.Cross, and P.C. Hohenberg, Rev. Mod. Phys.
{\bf 65}, 851 (1993).
\bibitem{JLB} Ch. Jung, M. L\"{u}cke, and P. B\"{u}chel, Phys. Rev. E
{\bf 54}, 1510 (1996).
\bibitem{LB}P. B\"{u}chel and M. L\"{u}cke, Phys. Rev. E
{\bf 61}, 3793 (2000).
\bibitem{RB}J. M. Luijkx, J. K. Platten, and J. C. Legros, Int. J.
Heat Mass Transfer {\bf 24}, 1287 (1981).
\bibitem{TC}K. L. Babcock, G. Ahlers, and D. S. Cannell, Phys.
Rev. E {\bf 50}, 3670 (1994).
\bibitem{n}The characteristic value of the velocity of the
trough-flow, which enters into the Reynolds number is that average
over the Poiseuille profile.
\bibitem{code}S. V. Patankar {\it Numerical Heat Transfer and
Fluid Flow\/} McGraw-Hill, 1980.
\bibitem{note}At the given values of the parameters the
linear stability analysis for the infinitely-extended in
$x$-direction layer yields the following thresholds $r^U \approx
1.12;\; r^D \approx 1.13$\cite{JLB}. In our simulations the
thresholds should be a little beyond these numbers due to the
finiteness of the aspect ratio and the stabilizing role of the
lateral boundary conditions ({\it size-effect}). Thus, for $r^U$ we
obtain 1.155. As for $r^D$, it is much more difficult to obtain its
value from the simulation due to the suppression of DTW by UTW (see
further discussion in the text). However, the good agreement
between $r^U$ obtained in our our simulation and in Ref.\cite{JLB}
allows us to suppose that $r^D$ in our case is also close to the
corresponding value of Ref.\cite{JLB}.
\bibitem{Cross}M. C. Cross, Phys. Rev. Lett. {\bf 57}, 2935
(1986).
\bibitem{note1}Precise quantitative comparison is difficult since
dependence $k^{U,D}_c(Re)$ in Ref.\cite{JLB} is given as a plot,
where two branches $k^U_c(Re)$ and $k^D_c(Re)$ merge at $Re=0.$
\bibitem{pure1}H. W. M\"{u}ller, and M. L\"{u}ke, Phys. Rev. A
{\bf 45}, 3714 (1992).
\bibitem{pure2}A. Couairon, and J. M. Chomaz, Phys. Rev. Lett.
{\bf 79}, 2666 (1997).
\end{references}
\end{document}